\documentclass[prb,twocolumn,showpacs]{revtex4}

\usepackage{graphicx}
\usepackage{multirow}
\usepackage{dcolumn}
\usepackage{bm}
\usepackage[normalem]{ulem}
\usepackage{amsmath}
\usepackage{amsfonts}
\usepackage{amssymb}
\usepackage{color}
\usepackage{colordvi}
\usepackage{dcolumn}

\begin{document}

\title{
Resistor model for the electrical transport
in quasi-one-dimensional organic (TMTSF)$_{2}$PF$_{6}$ superconductors under pressure
}

\date{\today}
\author{
H. Meier$^{1,3}$, P. Auban-Senzier$^{2}$, C. P\'epin$^{3}$, and D. J\'erome$^{2}$
}
\affiliation{
$^1$Institut f\"ur Theoretische Physik III, Ruhr-Universit\"at Bochum, 44780 Bochum, Germany\\
$^2$Laboratoire de Physique des Solides, 
Universit\'e Paris-Sud, 91405 Orsay, France\\
$^3$IPhT, CEA-Saclay, 
L'Orme des Merisiers, 91191 Gif-sur-Yvette, France
}

\begin{abstract}
  We  investigate the metallic transport in the organic superconductor (TMTSF)$_2$PF$_6$ under pressure within the framework of the spin density wave theory in the proximity of a Peierls quantum critical point (QCP). We use a simple transport model of hot and cold regions around the Fermi surface, driven by the proximity to the QCP in order to provide a template to fit the experimental results. Successful agreement with the data comforts the interpretation of extended criticality around the antiferromagnetic QCP in the normal phase of quasi one-dimensional organic superconductors.
\end{abstract}

\maketitle

%
%

\section{Introduction}

Anomalous  transport properties have been reported in the normal  phase of  the organic superconductors  (TMTSF)$_2$PF$_6$   and (TMTSF)$_2$ClO$_4$, also called Bechgaard salts \cite{DenisJerome1,bechgaard1},  as a function of temperature and applied pressure.  These quasi one-dimensional materials are remarkable in many respects, the least of all being the striking similarity of their temperature-pressure ($T\!-\!P$) phase diagram \cite{Jerome91} with other exotic superconductors, namely the iron-based superconductors \cite{doiron-leyraudSC,Jiang09}.   At ambient pressure, (TMTSF)$_2$PF$_6$ undergoes a magnetic phase transition of spin density wave (SDW) character, where the Fermi surface is destabilized by the onset of a spin density wave of the itinerant conduction electrons \cite{Jerome82,Yamaji98,Montambaux88}.
Under application of high pressure, the compound becomes superconducting with a maximum critical temperature~$T_c \approx 1.2\ \mathrm{K}$ \cite{DenisJerome1}.
Superconductivity coexists with the SDW order until the pressure reaches a critical value~$P_c$ above which the SDW order is no longer observed \cite{brazovskiijerome},
indicating the existence of an SDW quantum critical point (QCP).
At larger pressure, superconductivity shows a smooth decrease of $T_c$ with pressure.
The similarity with pnictide superconductors is not only seen in the ($T\!-\!P$) phase diagram, showing the simultaneous occurrence of SDW order and superconductivity, but also in the properties of the normal phase \cite{doiron-leyraudSC}.  Indeed, the electrical resistivity~$\rho$ shows a striking linear temperature dependence, $\rho-\rho_0 \propto T$, at~$11.8~\mathrm{kbar}$ and larger pressure in (TMTSF)$_2$PF$_6$, suggesting an unusual scattering mechanism that is possibly similar to the one present in the pnictides (and cuprates).

Extensive theoretical investigation has been carried out, based on a one-loop renormalization group (RG) formalism \cite{sedeki-bourbonnais} especially suitable for one-dimensional systems \cite{Giamarchi08a}. These studies have been reported to reproduce many of the most striking features observed experimentally \cite{sedeki-bourbonnais,efetovinc}.  Of special importance is the interplay between itinerant antiferromagnetism (SDW) and superconductivity, which controls the magnitude of the coupling constants in the vicinity of the QCP. An extensive regime of criticality has been uncovered, for which linear in $T$ resistivity is obtained down to lowest temperatures (within the one-dimensional framework) on a finite range of pressure. Furthermore, the $T$-linear  behavior disappears above an upper scale $T_0$ where the quadratic law is recovered.
Above $P_c$, the system is thought to be homogeneous, and although the quantum phase transition is reported to be weakly first-order \cite{brazovskiijerome}, it is believed that considerable quantum fluctuations are present to cause anomalous regimes, in both transport and thermodynamics.

In this paper, we build upon the RG insight and suggest an explicit model based on a Peierls instability in a \emph{quasi-one-dimensional system} driven to zero-temperature by the presence of warping on the Fermi surface. The idea  of driving a finite temperature Peierls transition towards a QCP end point through increasing the warping has been used in past studies of chromium \cite{ricechromium,chromiumpepinnorman}, a system showing almost perfect nesting between Fermi pockets via translation of the SDW wave vector. Here, we shall employ a model of the same type, where the SDW transition is driven to zero-temperature through the increase of the unnested  part of the electronic dispersion \cite{millis,Moriya,hertz}. To the best of our knowledge, the theoretical investigation of a Peierls-type QCP driven by curvature effects is unprecedented while the family of organic superconductors provides a suitable model system to check the theory.

%
%

\section{Model for a nesting QCP}

\begin{figure}[tbp]
\centerline{\includegraphics[width=\linewidth]{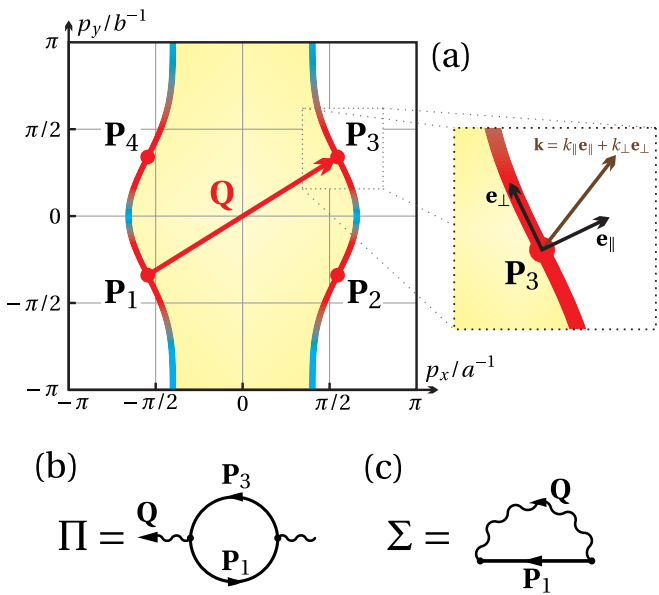}}
    \caption{(Color online) (a) Brillouin zone with the Fermi surface and the nesting vector~$\mathbf{Q}$ connecting two inflection points~$\mathbf{P}_1$ and~$\mathbf{P}_3$. Inset: Unit vectors~$\mathbf{e}_\parallel$ and~$\mathbf{e}_\perp$ parallel and perpendicular to the Fermi velocity at~$\mathbf{P}_3$ that define the coordinates~$\mathbf{k}=(k_\parallel,k_\perp)$.
     (b) Polarization bubble for the paramagnon mode. (c) Fermion self-energy. The wavy line represents
     the effective paramagnon from Eq.~(\ref{eqn1a}).}
\label{fig1}
\end{figure}

In the Bechgaard salts, the kinetic electronic spectrum is suitably modeled by the orthorhombic dispersion relation, \begin{align}
\label{eqn0}
\varepsilon(\bold{p}) = v_0(|p_x| - p_F) - t_b\cos(b p_y)-t_b'\cos(2b p_y)\ .
\end{align}
The corresponding Fermi surface is shown in Fig.~\ref{fig1}(a).
Generally $t_b'\ll t_b \ll v_0 p_F$ and the $t_b'$-unnesting term drives the system into criticality. For $t_b'=0$, the Fermi surface is perfectly nested with the nesting vector~$\bold{Q}=(2p_F,\pi/b)$ and at sufficiently low temperature the system is found to be in the SDW phase. For finite yet still small~$t_b'$, nesting is no longer perfect but still very good in the proximity of the inflection points --- the \emph{hot spots} --- of the Fermi surface. Eventually as $t_b'$ is increased further (by applying greater external pressure) beyond a critical value, nesting will be ineffective and the SDW order is destroyed.  Nesting of finite parts of the Fermi surface at high temperature is now reduced to the four inflection points~$\mathbf{P}_{1-4}$ at $T=0$ as depicted in Fig.~\ref{fig1}(a). This picture survives down to zero-temperature, implying a quantum phase transition at a critical value of the  coupling constant~$t_b'$.  Using the parameters of Ref.~\onlinecite{sedeki-bourbonnais}, we have $t_b$$ \sim 
200\ \mathrm{K}$, and the critical coupling $t_b' \approx 25.4\ \mathrm{K}$.

In order to study the physics around the hot spots, we expand the spectrum~(\ref{eqn0})
in their vicinity. Around~$\mathbf{P}_3$ and in leading order in~$t_b'$,
\begin{align}
 \varepsilon(\mathbf{P}_3+\mathbf{k}) \simeq v k_\parallel - b_3 k_\perp^3 - b_4 k_\perp^4\ .
    \label{eqn0a}
\end{align}
Herein, $k_\parallel$ and $k_\perp$ are the momentum components parallel and perpendicular, respectively, to the Fermi velocity at~$\mathbf{P}_3$, cf. the inset of Fig.~\ref{fig1}(a). The parameters in the reduced spectrum~(\ref{eqn0a}) are $v=v_0/\gamma$, $b_3 = b^3\gamma^3 t_b/6$, and $b_4 = - b^4\gamma^4 t_b'/2$ with $\gamma^{-1} = \sqrt{1+(bt_b/v)^2}$. Close to the ``nesting partner''~$\mathbf{P}_1$, the spectrum is similarly approximated as
\begin{align}
 \varepsilon(\mathbf{P}_1+\mathbf{k}) = \varepsilon(\mathbf{P}_3 - \mathbf{k}) \simeq -v k_\parallel + b_3 k_\perp^3 - b_4 k_\perp^4
 \ .   \label{eqn0b}
\end{align}
Note that we have expanded the spectrum up to the term of forth order in~$k_\perp$. Had we kept only terms until order~$k_\perp^3$, the reduced hot-spot spectra~(\ref{eqn0a}) and~(\ref{eqn0b}) would not contain
any mechanism to violate perfect nesting. Indeed, the simple coordinate transformation $k_\parallel \mapsto k_\parallel + v^{-1}b_3k_\perp^3$ illustrates that there would effectively be no curvature effects in the resulting physics so that, inevitably, we
would find a finite-temperature Peierls instability towards SDW for any externally applied pressure. In contrast, the quartic-order term clearly breaks nesting and, since $b_4\propto t_b'$, is an immediate consequence of the presence of the pressure-driven $t_b'$-warping term in Eq.~(\ref{eqn0}). With the range of transverse momenta~$k_\perp$ approximately between~$-\pi/(2b)$ and~$\pi/(2b)$, we find that the forth-order term becomes important as soon as~$t_b'/t_b \sim 2/(3\pi)$, which is fairly compatible with the critical value for~$t_b'$ given in Ref.~\onlinecite{sedeki-bourbonnais}. Increasing~$t_b'$ beyond its critical value reduces the Fermi surface
region for which the reduced spectra~(\ref{eqn0a}) and~(\ref{eqn0b}) are valid and thus sets a limit to the volume of the hot spots.

Looking for nontrivial effects such as the observed QCP, we need to enrich the model of noninteracting fermions by a proper model for the two-particle interaction. Previous works established that there are three relevant interaction channels in the Bechgaard salts: backward scattering (with the coupling constant~$g_1$), forward scattering ($g_2$), and Umklapp scattering ($g_3$) \cite{sedeki-bourbonnais}. The RG studies have shown that the superconducting fluctuations lead to a drastic decrease of the coupling constants~$g_1$ and~$g_2$, due to the interplay of Cooper and  SDW fluctuations at the nesting points, whereas the Umklapp coupling constant~$g_3$ remains essentially unaffected. For the following QCP study, we represent the fermion-fermion interaction as mediated by a bosonic mode that becomes critical at the QCP. Following the insight of Ref.~\onlinecite{sedeki-bourbonnais}, we retain only the coupling constant $g_3$ as medium for the electron-paramagnon interaction. Our approach allows us to analyze 
both thermodynamic and transport properties, but considering that $g_3$ is related to the Umklapp processes enables us to focus mainly on the resistivity behavior. We perform calculations neglecting vertex corrections to the one-loop self-energies ---
an approximation that we may expect to yield at least qualitatively the correct physical picture.

%
%

In a phenomenological low-energy picture, we may assume that after integrating out
all high-energy degrees of freedom, the effective interaction is mediated by long-wave
paramagnon modes. Here, we consider such a bosonic mode that transfers
a momentum of order~$\bold{Q}$, cf. Fig.~\ref{fig1}(a).
The coupling of the bosonic modes to the electrons generates a self-energy term~$\Pi_{\omega,\mathbf{q}}$ in the boson propagator~$\chi(\mathrm{i}\omega,\bold{Q}+\bold{q})$. In
the one-loop approximation, $\Pi_{\omega,\mathbf{q}}$ is given by the polarization bubble [see Fig.~\ref{fig1}(b)], whose relevant
nonanalytic part is
\begin{align}
\label{eqn1}
\Pi_{\omega,\bold{q}} &= \frac{g_3|q_\perp|}{4\pi^2v}\ \ln\bigg\{
 \frac{(2b_4q_\perp^4)^2}{\omega^2+ \xi_{\mathbf{q}}^2}
 \bigg\}
\end{align}
with $\xi_{\mathbf{q}}= vq_\parallel - b_3 q_\perp^3 + b_4 q_\perp^4$. Formula~(\ref{eqn1})
for the bosonic self-energy is valid for~$t_b'$ larger than the critical value~$\approx 25.4\ \mathrm{K}$, ensuring that that the unnesting terms $b_4k_\perp^4$ in Eqs.~(\ref{eqn0a})--(\ref{eqn0b}) are active.
It is important to note that the presence of the~$k_\perp^4$-term prevents the polarization from producing mass terms containing logarithms in temperature and thus establishes
the existence of a QCP --- for subcritical values~$t_b' \lesssim 25.4\ \mathrm{K}$, the effective absence of forth-order $k_\perp^4$-terms does lead to logarithmic $\ln T$-terms in the bosonic mass so that the phase transition towards an SDW state sets in at a finite temperature~$>0$. As the remaining analytic part of the bosonic spectrum is generically an analytic function of~$\xi_{\mathbf{q}}^2$, we thus write the effective propagator for the para\-magnons as
\begin{align}
\label{eqn1a}
\chi(\mathrm{i}\omega,\bold{Q}+\bold{q}) &= \frac{1}{\mu + \alpha (\xi_{\mathbf{q}}/v)^2 + \Pi_{\omega,\bold{q}}}
\end{align}
with~$\alpha\sim 1$. The bosonic mass~$\mu$ [In this work, we consider $\mu>0$.] measures the distance to the QCP and, close to it, we should consider the limit~$\mu \rightarrow 0$. The logarithm present in the paramagnon propagator~(\ref{eqn1a}) is characteristic of a Peierls phase transition and all the anomalous behavior shall ultimately be due to this nonanalyticity.

\section{Conductivity}

Using the effective model for the hot-spot electrons coupled to critical paramagnons, we are in a position to investigate their transport properties. The relevant quantity is the retarded electron self-energy~$\Sigma^{R}(\varepsilon)$, which within the precision of the one-loop approximation is represented by the Feynman diagram in Fig.~\ref{fig1}(c).  In the standard way for an itinerant electron QCP, the momentum dependence of the self-energy is negligible compared to the energy dependence
and the Matsubara self-energy is thus given by
\begin{align}
\Sigma(\mathrm{i}\varepsilon) = - g_3 T\sum_{\mathrm{i}\omega} \int \frac{\mathrm{d}\xi_\mathbf{q}}{2\pi v}
 \frac{\bar{\chi}(\mathrm{i}\omega,\xi_\mathbf{q})}{\mathrm{i}(\varepsilon+\omega)-\xi_\mathbf{q}}
\end{align}
with $\bar{\chi}(\mathrm{i}\omega,\xi_\mathbf{q}) = \int(\mathrm{d}q_\perp/2\pi)\ \chi(\mathrm{i}\omega,\bold{Q}+\bold{q})\big|_{\xi_\mathbf{q}=\mathrm{const}}$. Performing the analytical continuation to real-time frequencies~$\varepsilon$, we obtain in the limit $\varepsilon\rightarrow 0$ for the imaginary part of the retarded self-energy in the electronic Green's functions the formula
\begin{align}
\mathrm{Im}\ \Sigma^{R}(\varepsilon) &\simeq  \pi T\ \frac{\ln\big(p_F^{-2}\mu+ \varepsilon^2/\varepsilon_F^2\big)}{\ln\big(\varepsilon^2/\varepsilon_F^2\big)}
\ .
\label{2a03}
\end{align}
It shows that at criticality, $\mu=0$, the self-energy of hot-spot electrons is linear in temperature, $\mathrm{Im} \Sigma^{R}_{\mathrm{QCP}}(\varepsilon\!\rightarrow\! 0) = \pi T$, and independent from the coupling constants. As a straightforward consequence, the resistivity of the hot-spot electrons in the compound would at arbitrarily low temperatures be linear in~$T$ as well. Away from the QCP, the finite bosonic mass $\mu$ suppresses for frequencies $|\varepsilon|\lesssim v\sqrt{\mu}$ the quantity~$\mathrm{Im} \Sigma^{R}(\varepsilon)$, which for~$\varepsilon\!\rightarrow\! 0$ tends logarithmically to zero. As a result, since for the conductivity essential frequencies~$\varepsilon$ are of order~$T$, a linear law for the temperature dependence of the hot-spot resistivity appears only above the critical temperature
\begin{align}
\label{2a04}
T_S \sim v \sqrt{\mu}\ .
\end{align}
At the QCP, clearly, $T_S=0$. Note that, since the limit~$T\gg T_S$ is essentially equivalent to the limit~$\mu\rightarrow 0$, the coefficient in front of~$T$ in the resistivity effectively does not depend
on the value of the bosonic mass~$\mu$, i.e. on the applied pressure that determines~$\mu$.

%
%

We turn now towards a dichotomic description of the transport properties \cite{hublina-rice} of the compound in terms of hot-spot and cold-spot regions on the Fermi surface, hereby providing a simple model upon which to test the experimental data of (TMTSF)$_2$PF$_6$ from Ref.~\onlinecite{doiron-leyraudSC}. The conductivity~$\sigma(T)$ is the sum of contributions from the entire Fermi surface. Treating separately the contributions from the hot spots (with the Fermi surface volume fraction~$v_h$) and those due to the cold regions (volume $1- v_h$), we write~$\sigma(T)$ as the sum
\begin{align}
\sigma(T)= \frac{v_h}{\rho_0 + \rho_{\mathrm{hot}}(T) } + \frac{1-v_h}{\rho_0 + \rho_{\mathrm{cold}}(T)}\ .
\label{dichotomy}
\end{align}
In circuit language, see the inset of Fig.~\ref{fig2}, this formula corresponds to the parallel arrangement of the resistances due to the hot and cold regions of the Fermi surface while each of the two resistances is viewed as a series of the residual resistance and a specific temperature-dependent one. The residual resistivity~$\rho_0$ is experimentally given by the $T \rightarrow 0 $ limit and is the result of elastic scattering processes. Guided by the preceding theoretical considerations, we specify in the following the form of the temperature-dependent resistivities~$\rho_{\mathrm{hot}/\mathrm{cold}}(T)$ and their underlying scattering processes in the hot and cold regions.

For the cold regions, we may for all temperatures assume the quadratic law $\rho_{\mathrm{cold}}(T) =  B T^2$ accounting for the electron-electron scattering processes typical of the metallic  behavior.
At a sufficiently high temperature $T > T_0$, the notion of cold and hot regions is irrelevant so that we may expect
the same law also in the hot regions, $\rho_{\mathrm{hot}}(T) = B T^2$. Lowering the temperature, we encounter at a temperature~$T_0$ the crossover into the quantum critical regime. Here, (Umklapp) scattering of hot conduction electrons through the quantum critical paramagnons leads according to the preceding analysis to a linear law~$\rho_{\mathrm{hot}}(T) = A T$, cf. Eq.~(\ref{2a03}). Below a second crossover temperature $T_S$, Eq.~(\ref{2a04}),
the linear resistivity is suppressed and one should again expect a Fermi-liquid like behavior,
$\rho_{\mathrm{hot}}(T)= C T^2$, though with an effective quasi-particle mass heavily renormalized by
the interaction with paramagnons close to criticality. At the QCP, $T_S=0$ so that the linear law for~$\rho_{\mathrm{hot}}(T)$ prevails down to zero temperature while at very high pressure, the differentiation between hot and cold regions is no longer valid so that we expect $C \rightarrow B$ and the critical window between~$T_S$ and $T_0$ to shrink to zero. Table~\ref{tbl: summary} summarizes the temperature laws for the three regimes.
\begin{table}[t]
\caption{\label{tbl: summary} Temperature dependencies of the resistivities of hot and cold electrons.}
\begin{tabular}{c|cc}
  \noalign{\smallskip}
  \hline\hline
   \multirow{2}{*}{temperature region}  & \multicolumn{1}{c|}{\multirow{2}{*}{$\rho_\mathrm{hot}(T)$}} & \multirow{2}{*}{$\rho_\mathrm{cold}(T)$} \\
   &\multicolumn{1}{c|}{ }&\\
 \hline
 \multirow{2}{*}{$T>T_0$} & \multicolumn{2}{c}{\multirow{2}{*}{$BT^2$}} \\
 &&\\ \cline{2-3}
 \multirow{2}{*}{$\qquad T_S<T<T_0\qquad$} &
 \multicolumn{1}{c|}{\multirow{2}{*}{$\qquad AT\qquad$}} & \multirow{2}{*}{$\qquad BT^2\qquad$} \\
 &\multicolumn{1}{c|}{ }&\\
 \multirow{2}{*}{$T<T_S$} & \multicolumn{1}{c|}{\multirow{2}{*}{$CT^2$}} & \multirow{2}{*}{$BT^2$} \\
 &\multicolumn{1}{c|}{ }&\\
 \hline\hline
\end{tabular}
\end{table}

%
%

\section{Comparison with experiments}

Our dichotomic conductivity model
suggests a three-step analysis of the transport data on (TMTSF)$_2$PF$_6$ \cite{doiron-leyraudSC}: In the first step, the coefficient~$B$ is fixed from the quadratic resistivity law $\rho_0 +B T^2$ at high temperatures ($\sim 30\ \mathrm{K}$). For the residual resistivity~$\rho_0$, the zero-temperature extrapolation of the experimental data, we have
used the same values as in Ref.~\onlinecite{doiron-leyraudSC}. Then, we extract
in the second step the critical regime at intermediate temperatures where a significant linear temperature contribution is observed. Fitting here the data to the formula~(\ref{dichotomy}) written for temperatures $T_S<T<T_0$, we find the coefficient~$A$ and the hot-spot volume fraction~$v_h$. Note that $B$ has already been fixed and thus is not a free fitting parameter.
At the same time, our theory predicts that~$A$ is pressure-independent. Thus once~$A$ is determined for one pressure, e.g. the one at which the linearity in~$T$ prevails down to lowest temperatures, the only remaining free fitting parameter is $v_h$.
In the final third step, we similarly use the low-temperature ($T<T_S$) form of Eq.~(\ref{dichotomy}) to determine the coefficient~$C$. Within the philosophy of the resistor model, $A$ and $C$ are constants as a function of temperature inside the temperature regime they appear in. This ensures that the regimes are properly defined according to Eq.~(\ref{dichotomy}) and Table~\ref{tbl: summary}. Theoretically, we may expect logarithmic corrections, see Eq. (\ref{2a03}), but when comparing with experiments, these are fairly approximated by constants. Finally, we determine the crossover temperatures~$T_S$ and~$T_0$ as the intersections of the fits found for each regime.

\begin{figure}[h]
\centerline{\includegraphics[width=0.8\linewidth]{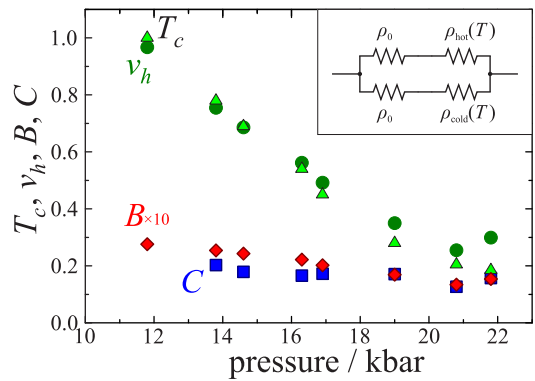}}
\caption{(Color online) Pressure dependence of the model parameters obtained from the fit: $10\times B$ (diamonds, in $\mu\Omega\ \mathrm{cm}/\mathrm{K}^2$), $C$ (squares, in $\mu\Omega\ \mathrm{cm}/\mathrm{K}^2$), and $v_h$ (circles) compared
to the temperature of the superconducting
transition~$T_c$ (triangles, in $\mathrm{K}$) from Ref.~\onlinecite{doiron-leyraudSC}. Inset: ``equivalent circuit'' for the dichotomic conductivity model~(\ref{dichotomy}).}
\label{fig2}
\end{figure}

Figure~\ref{fig2} shows the pressure dependence of~$B$, $C$, and $v_h$ as a result
of the analysis of $\sigma(T)$ between $0.15$ and $34\ \mathrm{K}$ at seven different pressures according to the three-step fitting procedure discussed above.
The analysis confirms that the contribution linear in~$T$ is indeed well-described by a pressure-independent coefficient~$A$ that if treated as a free fitting parameter mildly jitters around $A=0.38\ \mathrm{\mu \Omega\ cm/K}$.
The coefficient $B$ is related to the effective mass in the cold regions,  $B\sim m_{\parallel}^{2}$.
Its slight decrease under pressure in Fig.~\ref{fig2} can be ascribed to the increase of the intermolecular in-chain overlap, possibly enhanced by correlations. The data are also in agreement with the pressure dependence of the spin susceptibility measured by NMR experiments \cite{Wzietek93}.
The coefficient $C$ describing the increase of the effective electron mass at hot spots is roughly constant but its size is ten times larger than the order of magnitude of~$B$. The fact that such an enhancement does not fade away under pressure indicates that even under $21.8\ \mathrm{kbar}$ the scattering off antiferromagnetic spin fluctuations is still very strong. Under a pressure of $11.8\ \mathrm{kbar}$, corresponding to the point closest to the QCP, no quadratic law could be observed down to the lowest temperature after superconductivity had been removed by the application of a small magnetic field along $c^{\star}$. This suggests that the quantum critical regime of linear resistivity extends down to temperatures very close to zero at this point.
The hot-spot volume~$v_h$, which close to the QCP is $v_h \approx 0.97$, is decreasing under pressure ($v_h \approx 0.30$ at $21.8\ \mathrm{kbar}$). This is in accordance with the intuitive physical picture that
the distance from the QCP enhances unnesting and thus reduces the effective size of the hot spots. Its value remarkably follows $T_c$, in agreement with earlier findings \cite{doiron-leyraudSC}.

\begin{figure}[h]
\centerline{\includegraphics[width=0.9\linewidth]{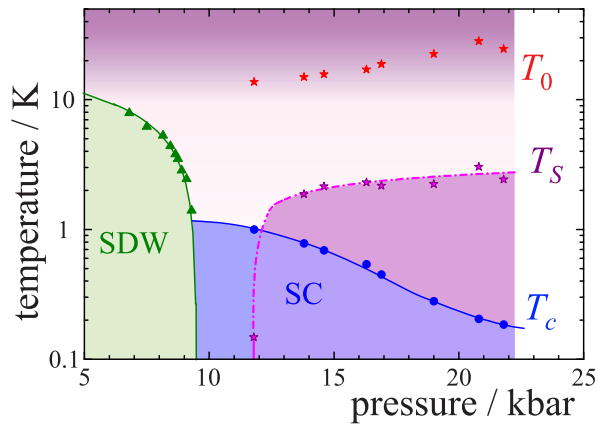}}
\caption{(Color online) (pressure$-$temperature) phase diagram of (TMTSF)$_2$PF$_6$ displaying
the crossover temperatures~$T_S$ and $T_0$ from this analysis as well as the long-range order phases. The lines are guides to the eye.
The transition temperatures towards~SDW (triangles) and~SC (circles) are those from Ref.~\onlinecite{doiron-leyraudSC}.
}
\label{fig3}
\end{figure}

Both crossovers  $T_0$  and  $T_S$ are plotted \textit{versus} pressure in Fig.~\ref{fig3}. The behavior of $T_S$ is strongly suppressed in the vicinity of the QCP, in fair agreement with Eq.~(\ref{2a04}). It is to be noted that while the hot-spot contribution to conductivity is dominant at the pressure of $11.8\ \mathrm{kbar}$ close to the QCP, the presence of the cold regions is crucial to explain the pressure decrease of the resistivity at a fixed temperature. Indeed, as discussed above, increasing the distance to the QCP induces a decrease of $v_h$, thus favoring the conduction through the cold regions at larger pressures. In the language of the equivalent circuit
(see inset of Fig.~\ref{fig2}), the less-resistive cold regions short-circuit the larger and for $T_S<T<T_0$ linear in~$T$ hot-spot resistance.

%
%

\section{Conclusion}

In conclusion, we present a theory of a QCP associated with the Peierls-type singularity. This theory is nontrivial as the role of the curvature is preponderant in stabilizing the logarithmic divergences and  yields strong influence on the form of the crossovers.  The  organic  Bechgaard salts  constitute an almost perfect  model system with the simplicity of their band structure allowing us to test the curvature effects.
Within a hot-spot/cold-spot dichotomic conductivity model for the itinerant electron QCP,
we confront the critical theory with the experimental data obtained in transport measurements for (TMTSF)$_2$PF$_6$,
showing a good agreement. At the hot spots, the physics of the Bechgaard salts shows strong similarity with the physics in heavy-fermion systems.

\section*{Acknowledgements}

We acknowledge a fruitful cooperation with the team at Sherbrooke where the data in Ref.~\onlinecite{doiron-leyraudSC} were	 obtained. We acknowledge very useful discussions with K. B. Efetov and S.S. Brazovskii. H. M. acknowledges financial support from the SFB/TR~12 of the Deutsche Forschungsgemeinschaft and is grateful for the Chaire Blaise Pascal Fellowship of K.B. Efetov, which enabled his extended visit to  the IPhT at the CEA-Saclay.

%
%

\end{document}